\documentclass[conference]{IEEEtran}
\IEEEoverridecommandlockouts
\usepackage{cite}
\usepackage{amsmath,amssymb,amsfonts}
\usepackage{graphicx}
\usepackage{textcomp}
\usepackage{multirow}
\usepackage{booktabs}
\usepackage{epstopdf}
\usepackage{algorithm}
\usepackage[algo2e]{algorithm2e} 
\usepackage{algpseudocode}
\usepackage{url}
\usepackage{hyperref}
\usepackage[utf8]{inputenc}
\usepackage[english]{babel}
\begin{document}
\title{\textbf{StationPlot}: A New Non-stationarity Quantification Tool for Detection of Epileptic Seizures*}
\author{\IEEEauthorblockN{Sawon Pratiher$^\S$, Subhankar Chattoraj$^\dagger$ and Rajdeep Mukherjee$^\ddagger$}
\IEEEauthorblockA{\textbf{$^\S$}\textit{Indian Institute of Technology Kharagpur, India} \\
\textbf{$^\dagger$}\textit{Techno India University, Salt Lake, India} \\
\textbf{$^\ddagger$}\textit{Manipal University, Jaipur, India}\\
\thanks{*This paper is accepted for presentation at IEEE Global Conference on Signal and Information Processing (IEEE GlobalSIP), California, USA, 2018}
}}
\maketitle
\thispagestyle{empty}
\pagestyle{empty}
\begin{abstract}
A novel non-stationarity visualization tool known as StationPlot is developed for deciphering the chaotic behavior of a dynamical time series. A family of analytic measures enumerating geometrical aspects of the non-stationarity \& degree of variability is formulated by convex hull geometry (CHG) on StationPlot. In the Euclidean space, both trend-stationary (TS) \& difference-stationary (DS) perturbations are comprehended by the asymmetric structure of StationPlot's region of interest (ROI). The proposed method is experimentally validated using EEG signals, where it comprehend the relative temporal evolution of neural dynamics \& its non-stationary morphology, thereby exemplifying its diagnostic competence for seizure activity (SA) detection. Experimental results \& analysis-of-Variance (ANOVA) on the extracted CHG features demonstrates better classification performances as compared to the existing shallow feature based state-of-the-art \& validates its efficacy as geometry-rich discriminative descriptors for signal processing applications.
\end{abstract}
\vspace{0.2cm}
\section{Introduction}
Robust biomarkers are essential for early stage epileptic seizure detection from electroencephalogram (EEG) signals. EEG signal contains information related to neurological activities of the human brain [1] \& neural impediments in the transmission flow paths manifests in chaos within the brain segments, resulting in physio-pathological brain-related disorders [2]. Several time-domain techniques such as wavelet-chaos in EEG sub-bands [1], empirical mode decomposition (EMD), EMD modulation bandwidths \& least-square support vector machine (LS-SVM) [2], EMD-variation features [3], Hilbert marginal spectrum analysis [4] \& Hilbert Huang Transform (HHT) aided field programmable gate array (FPGA) [5] implementation have been reported for SA detection. Further, joint time-frequency techniques like wavelet transform \& machine learning classifiers [6, 7], intra-cranial spatiotemporal correlation structures [8], time-frequency image descriptors [9] \& multi-level wavelet decomposition [10] have shown potency in epileptic seizure diagnosis. Non-linear methods like permutation entropy \& support vector machines (SVM) [11], multi-scale sample entropy [12], rational discrete short-time Fourier transform [13], sparse coding using Slantlet transform [14] for multi-resolution quantification \& sparse representation of seizure free (interictal) \& seizure (ictal) EEG signals is studied in [15]. Application of local pattern transformation features [16], ARIMA-GARCH models [17] \& $L_{l}$ regularized EEG attributes [18] have shown adequacy as potential SA biomarkers. Mode decomposition based EEG modeling [2-4] \& multi-fractal detrended fluctuation analysis (MFDFA) [23, 24] have been studied to exploit the underlying non-linearity \& non-stationary nature of EEG fluctuations. High-dimensional EEG data embedding via t-SNE based projection by maximizing the inter-class \& minimizing the intra-class variances of seizure \& healthy class with high probability of visualization is discussed in [25]. A comprehensive feature engineering for EEG signal analysis can be traced from [1, 2, 21], where the significance of an exhaustive \& robust geometry-rich model to abstract the non-stationary EEG time series analytics cannot be over-emphasized. As such, a new non-stationarity quantification tool, \textit{a.k.a,} StationPlot is proposed. Stationplot is a novel geometry rich non-stationarity quantification tool for generic time series analysis. It gives succinct feature engineering \& analytical insights about the temporal evolution of stationarity in a time series. \\ Ongoing, StationPlot theory \& its CHG features are given in Section II. Experimental results on EEG dataset is given in Section III. Section IV concludes the paper.
\section{Related Theory \& Method}
\subsection{Trend-stationary (TS) \& Difference stationary (DS) Signals}
Non-stationary signals have an unpredictable joint probability distribution due to combined presence of trends, random walks \& cycles, which poses significant difficulty in statistical parameter estimation such as mean \& co-variances. De-trending or successive differencing of a non-stationary signal yields a TS or DS time-series respectively [19]. Deterministic mean trend removal of a signal results in persistent forecasting interval widths, which exhibits recurrent long-run trending behavior, whereas successive differencing is required for stochastic mean trend removal. If a time-series becomes stationary after differencing \textit{'D'} times, it is called \textit{D$^{th}$} order DS time-series. Such signals never recuperate from the random fluctuations \& forecasting intervals grows over time [19]. Mathematically, Box-Jenkins non-stationary time series analysis by successive differencing to attain stationarity can be traced as $\Delta ^{D_{yt}}=\mu+\Psi (L)\varepsilon _t $. Where, $\Delta ^{D}=(1-L)^D$ is the $D^{th}$ order differencing operator \& $\Psi(L)=(1+\Psi_1L+\Psi_2L^2+...)$ is the infinite-degree operator polynomial with coefficients being absolutely summable \& all roots lying outside the unit circle.
\begin{figure}[!ht]
\centering
\includegraphics[width=9cm, height=7cm]{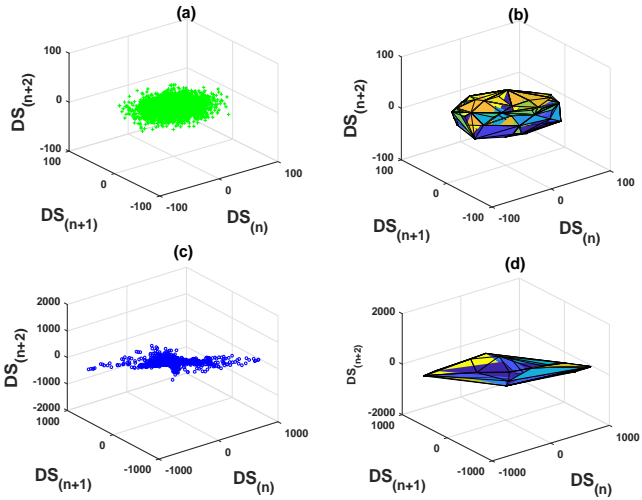}
\caption{Representative (a) Healthy EEG signal (in green), (b) 3-D StationPlot of (a), (c) Seizure EEG signal (in blue), (d) 3-D StationPlot of (c)}
\label{Fig.1}
\end{figure}
\subsection{StationPlot}
The trend present in the given time series, $x_t$ is de-trended by subtracting the mean value or linear trend of the feature vector \& a least-squares fit of the time series is envisaged. For the original time series, we have, $x_t=\mu_t+\varepsilon_t$. Where, $\mu_t$ signifies the deterministic mean trend in the time series \& $\varepsilon_t$ denotes the stochastic stationary process with zero means. De-trending is effectuated by parametric regression techniques or by non-parametric decomposition using filters such that: $X_2t\approx \varepsilon_t$. In the Euclidean space, we define the 2D \& 3D StationPlot \& its subsequent feature extraction thereof. For, the 2D-planar case, $n^{th}$ order StationPlot is defined as the plot of $X_1(n)$ versus $X_2(n)$, where, 
\begin{equation}
    X_1(n)= \Delta^{n}X(t)\ \&\ X_2(n)= \Delta^{{(n+1)}}X(t)
\end{equation}
While, for 3D, $n^{th}$ order StationPlot is defined as the surface generated by plotting $X_1(n)$, $X_2(n)$ \& $X_3(n)$ along the $X_1$, $X_2$ \& z-axis in $R^{3}$. s.t.,
\begin{multline}
    X_1(n) = \Delta^{n}X(t), X_2(n)= \Delta^{{(n+1)}}X(t)\   \&   \\X_3(n)= \Delta^{{(n+2)}}X(t)\ \&,\\ \Delta^{n}X(t)=\Delta^{n-1}X(t)-\Delta^{n-1}X(t-1)
\end{multline}
2D-StationPlot depicts the analytical variability of $\Delta^{n}X(t)$ versus \begin{math}\Delta^{{(n+1)}X(t)}\end{math}. It indicates the successive differencing effects on data sampling with system dynamics approaching continuum in the limiting scenario of sampling. In order to quantify the variability \& asymmetric behavior of the StationPlot, convex hull analytics (CHA) of the sampled data-points is done for variability characterization. Fig. 1 shows typical representative 3D-StationPlot of EEG signals.
\begin{algorithm}
\SetAlgoLined
\caption{\textbf{:} 2-D StationPlot visualization}
\textbf{Input:} (Degree of Trend-removal) / (Differencing) order = m,n \\ $f(n)\cong$ EEG time series, $\forall n\epsilon$ EEG data points\\
\textbf{Output:} 2D-stationPlot representation of EEG signal. 
\begin{algorithmic}[1]
\State function 2D-StationPlot ($f(n)$, m, n)
\State \textbf{Local variables:} $X_1$, $X_2$, $n_{max}=$ size (Data points set)
\While{$i<n_{max}$}{
 \Statex $X_1(i)\gets$ TS/DS [$f_n$, order m] 
 \Statex $X_2(i)\gets$ TS/DS [$f_n$, order n]
}
\State Plot $X_1$ \& $X_2$, on the 2D euclidean space
\end{algorithmic}
\end{algorithm}
\subsection{Convex hull geometry (CHG) on StationPlot}
\subsubsection{Convex hull (CH) of a set X}
In the affine space, CH defines the convex envelope which completely encloses all the data points in X, such that no concavities are present within it [21]. The convex combination of \textit{k} number of data-points in X, i.e., $x_1, x_2,..., x_k $ \& with \textit{k} number of constraints $\theta_1, \theta_2,...\theta_k\geq0 $ \& $\theta_1+ \theta_2+...+\theta_k=1$ is defined as: 
\begin{equation}
    x=\theta_1x_1+ \theta_2x_2+\theta_kx_k
\end{equation}
CH spans the set of all possible convex combinations of data-points in X by taking all the permutation of the coefficients, $\theta_k$. In the closed form, the convex hull can be expressed as: 
\begin{equation}
    conv(s)=\left \{\sum ^{|s|}_{i=1} \theta_ix_i| (\forall i:\theta_i\geq 0)\cap \sum ^{|s|}_{i=1}\theta_i=1 \right \}
\end{equation}
For our case, the set X denotes all the time series data-points projected in the n-Dimensional StationPlot. In the present study, n is restricted to 2 \& 3. Quickhull algorithm [20] has been used to compute the CH \& CHG thereof. Following CHG of the StationPlot, various features pertaining to the non-stationarity space geometry analytical significance are comprehended. Fig. 2 shows a representative 2-D StationPlot.
\begin{figure}[!ht]
\centering
\includegraphics[width=9cm,height=4cm]{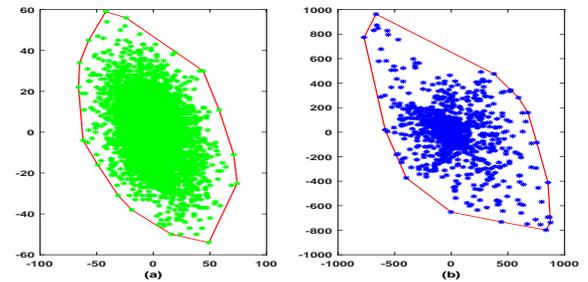}
\caption{Representative 2-D StationPlot of (a) Healthy (in green) \& (b) Seizure (in blue) EEG signals.}
\label{Fig.2}
\end{figure}
\begin{figure*}[!ht]
\centering
\includegraphics[width=18cm, height=3.5cm]{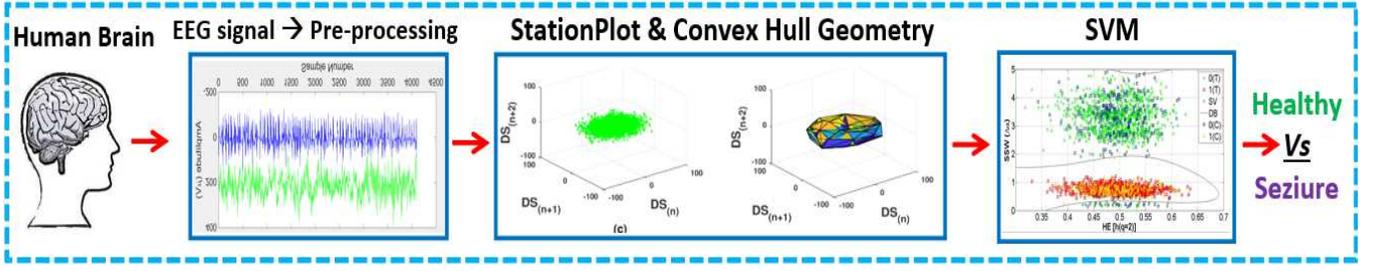}
\caption{Pipeline for proposed StationPlot based seizure detection.}
\label{Fig.3}
\end{figure*}
\subsection{Extraction of CHG features from StationPlot}
\subsubsection{Convex hull area/volume (CHA/V)}
CHA/V quantifies the polygon area/volume formed by the CH triangulation of the boundary of the CH \& signifies the total spread of the ROI on the StationPlot. For n points, $x_i,y_i$ lying on the convex hull in $R^2$, CH area is given by, 
\begin{equation}
    CHA=\frac{1}{2}\left ( \begin{vmatrix}
x_1 & x_2\\ 
y_1 & y_2
\end{vmatrix}+\begin{vmatrix}
x_2 & x_3\\ 
y_2 & y_3
\end{vmatrix}+...+ \begin{vmatrix}
x_n & x_1\\ 
y_n & y_1
\end{vmatrix}\right )
\end{equation}
Where, $|M|$ represents the determinant \& $i=1,2,3,...n$ denotes \textit{'n'} number of convex hull data points. Similarly, CHV represents the volume of the 3-simplex in the 3D space. 
CH $area/volume$ is a measure of the chaos strength \& degree of non-statistical fluctuations present in EEG signals.
\subsubsection{Convex Hull Perimeter (CHP)}
CHP is defined as the length of the CH circumference boundary or the aggregate path length of all data point's in the convex combination. It is computed by addition of all the adjoining vertices taken sequentially of the convex hull. If $(x_i,y_i)$ \& are $(x_{i+1},y_{i+1})$ the 2 adjacent vertices of the convex hull. Distance between them is given by $\sqrt{{(x_i-x_{i+1})}^2-{(y_i-y_{i+1})}^2}$. CHP is mathematically written as:
\begin{equation}
    CHP=\sum ^n_{i=1}=(\sqrt{{(x_i-x_{i+1})}^2-{(y_i-y_{i+1})}^2})
\end{equation}
\subsubsection{Circularity (C)}
The perturbed elliptical shape of the 2D-StationPlot's ROI exhibits asymmetric geometry, which is apprehended by \textit{C} measuring the degree of roundness of the convex hull \& its deviation from its circular nature. \textit{C} is independent of linear transformations. It is defined as:
\begin{equation}
    C=\frac{4*(CHA)*\pi}{{CH}^2}
\end{equation}
\subsubsection{Aspect ratio}
The aspect ratio measures the ratio of ROI’s main inertia axis length, $I_{main}$ to ROI’s minor inertia axis length, $I_{minor}$. 
\begin{equation}
    Aspect\ ratio=\frac{I_{main}}{I_{minor}}
\end{equation}
\section{Experimental Results \& Discussion}
\subsection{EEG dataset \& its acquisition}
EEG signals being highly non-linear \& non-stationary in nature has been used for experimental validation of the proposed method was validated using. We have used publicly available EEG dataset provided by Bonn University [26]. It comprises of five subsets designated as ‘A’, ‘B’, ‘C’, ‘D’ \& ‘E’, with each containing 100 records of 4097 data points \& 23.6s duration. Rudimentary details can be traced from [26]. The samples were digitized utilizing a 12-bit analog-to-digital converter \& with a sampling frequency of 173.61 Hz. Subsets E contains seizure activity while subsets ‘C’ \& ‘D’ are taken from five patients in the seizure-free interval \& recorded from the hippocampal formation region from the opposite hemisphere of the brain enclosing the epileptogenic zone. Subsets ‘A’ \& ‘B’ comprises of surface EEG recordings from five healthy individuals \& acquired by following the standard 10-20 electrode placement scheme. In the present study two classification problem, namely \textit{ABCD vs E} \& \textit{A vs E} are considered for non-stationarity analysis \& classification thereof using the developed tool.
\begin{figure}[!ht]
\centering
\includegraphics[width=9cm]{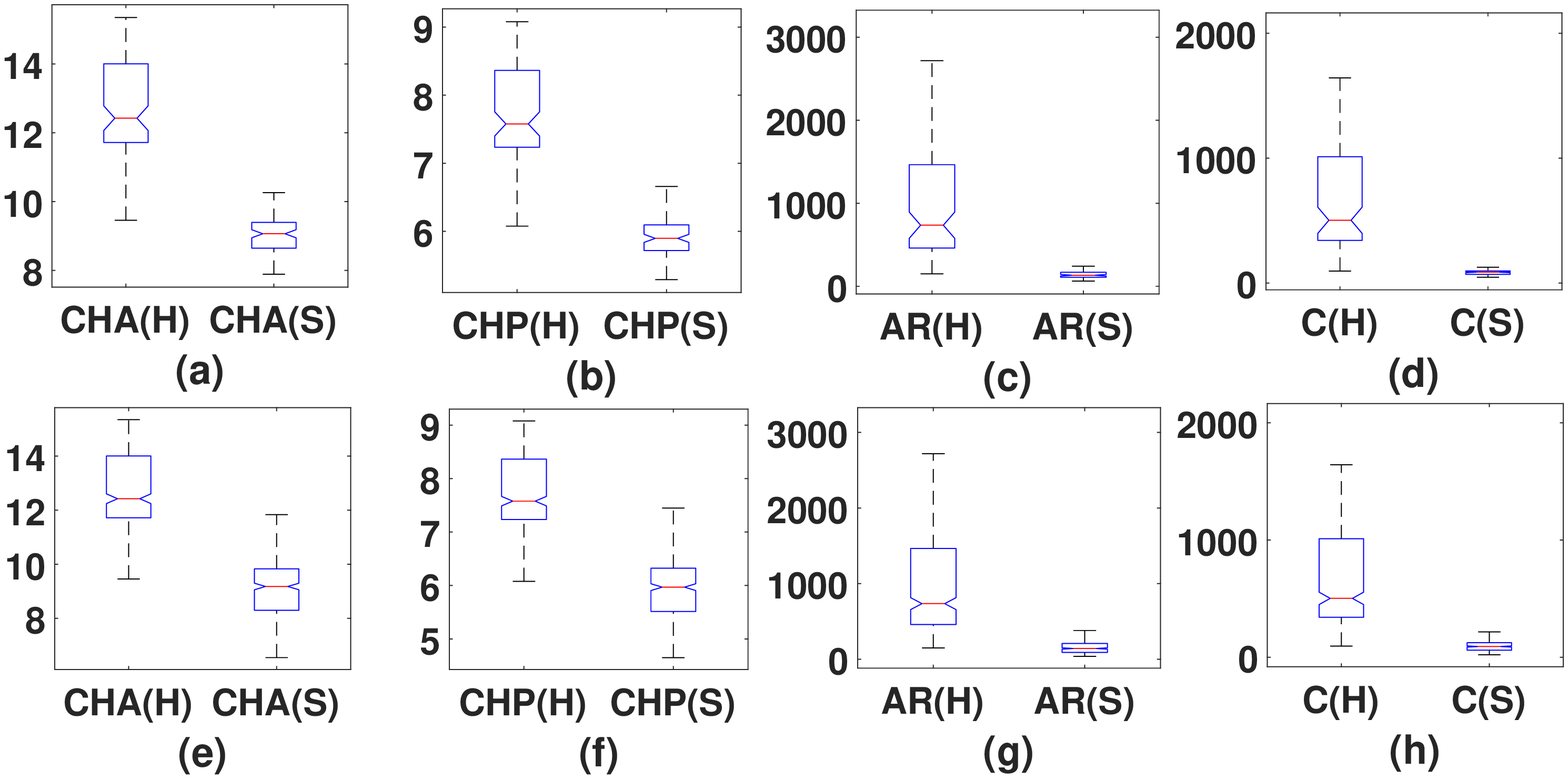}
\caption{Box-Plot of the extracted CHG features: (a)-(d) for (E vs ABCD) problem \& (a)-(d) for (E vs A) problem. H=Healthy \& S=Seizure}
\label{Fig.4}
\end{figure}
\subsection{Statistical analysis}
Prior to data sampling, the obtained EEG data records are subjected to band-pass filtering based pre-processing. For each of these 5 classes, CHG features are extracted from StationPlot's ROI \& are initially subjected to Analysis-of-variance (ANOVA) test for testing p-values. ANOVA, a statistical test used to analyze the group means differences in a sample \& characterize the discriminating significance in terms of p-values. A smaller p-value is associated with a stronger evidence in favor of the alternative hypothesis that the feature is effective for segregation. Fig. 4 shows box-plot of the extracted convex hull geometric feature variations. It clearly exhibits that the non-overlapping distribution band of the feature values. Table I gives the corresponding p-values of the feature descriptors for the two-class classification problem as contemplated \& demonstrates the class discriminating capability of the proposed CHG feature attributes. It can be noted that p-values of $\leq 0.01$ from Table 1 signifies the efficacy of CHG features in SA detection using EEG signals. 
\subsection{Classification Methodology \& Performance Measures}
By employing Kruskal-Wallis based feature ranking, most dominant features are selected \& fed to support vector machine (SVM) classifier for classification. In the present study, we confine ourselves to the following kernels: linear, quadratic, polynomial \& RBF kernel, where the optimal kernel parameters selection are governed by Mercer's theorem. For each of these feature vector classes, 70\% samples are randomly selected for training \& the rest 30\% is used for testing. These data partitioning, training \& testing, is repeated 100 times with training \& testing samples randomly selected for each run \& the mean performance measures are reported. Following measures are used: sensitivity (SN), specificity (SP), \& accuracy (AC) are defined as:
$SN=\frac{TP}{TP+FN}$, $SP=\frac{TN}{TN+FP}$ \& $AC=\frac{TP+TN}{TP+TN+FP+FN}$. 
Where, TP = true positive, TN = true negative, FP = false positive \& FN = false negative respectively.
\begin{table}[!ht]
\centering
\caption{p-Values of the features extracted @ 1\% significance level.}
\begin{tabular}{@{}|l|c|c|@{}}
\toprule
\multicolumn{1}{|c|}{\multirow{2}{*}{\textbf{Features}}} & \multicolumn{2}{c|}{\textbf{p-Values}} \\ \cmidrule(l){2-3} 
\multicolumn{1}{|c|}{} & \multicolumn{1}{c|}{\textbf{A vs E}} & \multicolumn{1}{c|}{\textbf{ABCD vs E}} \\ \midrule
Convex hull geometry (CHG) & $6.75 \times 10^{-16}$ & $4.68 \times 10^{-14}$ \\ \midrule
Convex Hull Perimeter (CHP) & $8.18 \times 10^{-15}$ & $3.19 \times 10^{-12}$ \\ \midrule
Circularity (C) & $6.11 \times 10^{-12}$ & $4.02 \times 10^{-10}$ \\ \midrule
Aspect ratio (AR) & \multicolumn{1}{l|}{$4.04 \times 10^{-7}$} & \multicolumn{1}{l|}{$1.05 \times 10^{-6}$} \\ \bottomrule
\end{tabular}
\end{table}
\begin{table}[!ht]
\centering
\caption{Performance Analysis of the Propose Method on the 2-Class problem. AC = accuracy, SE = sensitivity \& SP = specificity.}
\begin{tabular}{@{}|c|c|c|c|@{}}
\toprule
\textbf{\begin{tabular}[c]{@{}c@{}}Kernel Function \end{tabular}} & \textbf{\begin{tabular}[c]{@{}c@{}}Evaluation \\ Measure (\%)\end{tabular}} & \textbf{\begin{tabular}[c]{@{}c@{}}A vs E\\ (Mean $\pm$ Std.)\end{tabular}} & \multicolumn{1}{l|}{\textbf{\begin{tabular}[c]{@{}c@{}}ABCD vs E\\ (Mean $\pm$ Std.)\end{tabular}}} \\ \midrule
\multirow{3}{*}{Linear} & AC & 99.31 $\pm$ 1.12 & 98.70 $\pm$ 1.68 \\ \cmidrule(l){2-4} 
 & SN & 99.67 $\pm$ 1.05 & \textbf{98.74 $\pm$ 2.10} \\ \cmidrule(l){2-4} 
 & SP & \textbf{97.91 $\pm$ 1.79} & 96.13 $\pm$ 4.12 \\ \midrule
\multirow{3}{*}{Quadratic} & AC & 99.16 $\pm$ 1.16 & 97.66 $\pm$ 4.57 \\ \cmidrule(l){2-4} 
 & SN & 99.46 $\pm$ 1.23 & 98.37 $\pm$ 2.19 \\ \cmidrule(l){2-4} 
 & SP & 96.92 $\pm$ 2.01 & 95.71 $\pm$ 6.12 \\ \midrule
\multirow{3}{*}{\begin{tabular}[c]{@{}c@{}}Polynomial\\ (order =3)\end{tabular}} & AC & 98.85 $\pm$ 1.69 & 97.46 $\pm$ 1.71 \\ \cmidrule(l){2-4} 
 & SN & 98.96 $\pm$ 2.58 & 98.15$\pm$ 2.24 \\ \cmidrule(l){2-4} 
 & SP & 97.16 $\pm$ 2.67 & \textbf{98.57 $\pm$ 2.13} \\ \midrule
\multirow{3}{*}{\begin{tabular}[c]{@{}c@{}}RBF \\ $(\sigma =2 )$ \end{tabular}} & AC & \textbf{99.63 $\pm$ 1.60} & \textbf{98.79 $\pm$ 1.66} \\ \cmidrule(l){2-4} 
 & SN & \textbf{100 $\pm$ 0.00} & 98.18 $\pm$ 1.81 \\ \cmidrule(l){2-4} 
 & SP & 97.35 $\pm$ 3.15 & 93.10 $\pm$ 5.48 \\ \bottomrule
\end{tabular}
\end{table}
\begin{table}[!ht]
\caption{Seizure detection results of the proposed \& state-of-art methods: two-class problem (A vs E), N.A.: Not available.}
\centering
\begin{tabular}{|p{0.15\linewidth}|p{0.42\linewidth}|p{0.05\linewidth}|p{0.05\linewidth}|p{0.05\linewidth}|}
\hline
\textbf{Ref,.YoP} & \textbf{Methodology + Classifier} & \multicolumn{3}{ c|}{\textbf{Performance(\%)}}\\ 
\cline{3-5}
\textbf{}& \textbf{}  & \textbf{AC} & \textbf{SN} & \textbf{SP} \\
\hline
    \begin{tabular}[c]{@{}l@{}}{[}6{]} (2012)\end{tabular} & \begin{tabular}[c]{@{}l@{}}Discrete wavelet transform \\(DWT), normalized coefficient of \\variation (NCOV), LDA.\end{tabular} & 91.8  & 83.6 &100\\ 
    \midrule     
    \begin{tabular}[c]{@{}l@{}}{[}11{]} (2012)\end{tabular} & \begin{tabular}[c]{@{}l@{}}Permutation Entropy (PE)\\ SVM.\end{tabular} & 93.8 & 94.3& 93.2\\ 
    \midrule
    \begin{tabular}[c]{@{}l@{}}{[}21{]} (2013)\end{tabular} & \begin{tabular}[c]{@{}l@{}}Lacunarity \& Bayesian linear \\discriminant analysis (BLDA)\end{tabular} & 96.6 & 96.2& 96.7\\ 
    \midrule
    \begin{tabular}[c]{@{}l@{}}{[}7{]} (2014)\end{tabular} & \begin{tabular}[c]{@{}l@{}}Discrete wavelet transform\\ (DWT), fractal dimension (FD),\\ SVM.\end{tabular} & 97.5 & 98.0& 96.0\\ 
    \midrule
    \begin{tabular}[c]{@{}l@{}}{[}22{]} (2016)\end{tabular} & \begin{tabular}[c]{@{}l@{}}Weighted-permutation entropy\\ (WPE), SVM.\end{tabular} & 97.2 & 94.5& 100\\ 
    \midrule
    \begin{tabular}[c]{@{}l@{}}{[}10{]} (2016)\end{tabular} & \begin{tabular}[c]{@{}l@{}}Multi-level Wavelet \\Decomposition, ELM.\end{tabular} & N.A. & 99.4& 77.1\\ 
    \midrule    
    \begin{tabular}[c]{@{}l@{}} \textbf{This work}\end{tabular} & \begin{tabular}[c]{@{}l@{}}\textbf{StationPlot, SVM}\end{tabular} & \textbf{99.6} & \textbf{100}& \textbf{97.9}\\
    \midrule
\end{tabular}
\end{table}
Table II demonstrates the performance comparison on 2-class classification problem (CP) with SVM classifier \& different kernels. It can be found that linear kernel performs significantly better as compared to the other kernels which shows the robust computationally inexpensive classifier design on the proposed feature vectors. An overall classification accuracy of 99.31\% for (A vs E) \& 98.79\% for (ABCD vs E) 2-class (CP) is obtained respectively. Table III \& IV higlights the state-of-the-art comparative evaluation, which demonstrate the superior performances of the proposed methodology. Seizure EEG manifests in larger magnitude of random fluctuations. Recently computational intensive deep learning methods like convolutional neural networks (CNN) \& recurrent neural networks (RNN) have shown efficacy, but it requires heuristic hyper-parameter tuning for optimal performance, whereas StationPlot is an analytical closed form non-stationarity quantification framework subjugate these drawbacks \& give state-of-the-art performances. 
\begin{table}[!ht]
\caption{Seizure detection results of the proposed \& state-of-art methods: two-class problem (ABCD vs E), N.A.: Not available.}
\centering
\begin{tabular}{|p{0.15\linewidth}|p{0.42\linewidth}|p{0.05\linewidth}|p{0.05\linewidth}|p{0.05\linewidth}|}
\hline
\textbf{Ref,.YoP} & \textbf{Methodology + Classifier} & \multicolumn{3}{ c|}{\textbf{Performance(\%)}}\\ 
\cline{3-5}
\textbf{}& \textbf{}  & \textbf{AC} & \textbf{SN} & \textbf{SP} \\
\hline
    \begin{tabular}[c]{@{}l@{}}{[}3{]} (2013)\end{tabular} & \begin{tabular}[c]{@{}l@{}}Empirical Mode
    Decomposition-\\Modified Peak Selection \\(EMD-MPS), KNN\end{tabular} & 98.2 & N.A. & N.A.\\
    \midrule
    \begin{tabular}[c]{@{}l@{}}{[}4{]} (2015)\end{tabular} & \begin{tabular}[c]{@{}l@{}}Hilbert marginal spectrum \\(HMS), SVM \end{tabular} & 98.8 & N.A. & N.A.\\
    \midrule
    \begin{tabular}[c]{@{}l@{}}{[}16{]} (2017)\end{tabular} & \begin{tabular}[c]{@{}l@{}}Local Neighbor Descriptive \\Pattern (LNDP), One-dimensional \\Local Gradient Pattern (1D-LGP), \\ANN\end{tabular} & 98.7 & 98.3& 98.8\\
    \midrule
    \begin{tabular}[c]{@{}l@{}} \textbf{This work}\end{tabular} & \begin{tabular}[c]{@{}l@{}}\textbf{StationPlot, SVM}\end{tabular} & \textbf{98.8} & \textbf{98.7}& \textbf{98.6}\\
    \midrule
\end{tabular}
\end{table}
\section{CONCLUSION}
A novel non-stationarity quantification tool for early stage SA detection using EEG signals is presented. The proposed method can be used as an effective chaos modeling tool for any non-stationary time series visualization \& analysis of temporal evolutionary behavior \& its quantification thereof, in terms of trend-stationary \& difference-stationary components via CHG. The $D^{th}$ order differencing statistics exhibits significant knowledge about the underlying system progression over time \& adequately captures the underlying non-stationarity structure analytics, which is otherwise inaccessible. Moreover, classification accuracy of 99.31\% for (A vs E) \& 98.79\% for (ABCD vs E) problem, shows its adequacy in EEG signal classification. Ongoing, we are escalating our method for noise robustness via inclusion of area moments to study the distribution of non-stationary points w.r.t. to any arbitrary axis.
\section*{Acknowledgment}
The authors would like to thank Ralph G. Andrzeja et.al. for making the EEG dataset publicly available. 
\addtolength{\textheight}{-7cm}   

\end{document}